\documentclass[prresearch,aps,twocolumn,superscriptaddress,notitlepage]{revtex4-2}
\usepackage{amssymb,epsfig,amsmath,bm,subfigure,graphicx,textcomp,url,float,color,cases}
\usepackage{epsfig}
\usepackage{textcomp}
\usepackage[normalem]{ulem}

\usepackage[colorlinks=true, urlcolor=blue, anchorcolor=blue, citecolor=blue,filecolor=blue,linkcolor=blue,menucolor=blue
]{hyperref}

\usepackage{color}

\usepackage[titletoc,title]{appendix}

\begin{document}
\title{Macroscopic localization and collective memory in Poisson renewal resetting}


\author{Ohad Vilk}
\email{ohad.vilk@mail.huji.ac.il}
\affiliation{Racah Institute of Physics, Hebrew University of Jerusalem, Jerusalem 91904, Israel}

\begin{abstract}
Stochastic renewal processes are ubiquitous across physics, biology, and the social sciences. Here, we show that continuous-time renewal dynamics can naturally produce a mixed discrete-continuous structure, with a macroscopic fraction of particles occupying a discrete state. For ensembles of continuous-time random walkers subject to Poissonian renewal resets, we develop an age-structured framework showing this discrete component corresponds to localization at the reset configuration. We next show that collective interactions can retain memory although all reset events are memoryless. Remarkably, the transition to collective memory is discontinuous, and we identify a discontinuous dynamical phase transition from weak collective bias, where the dynamics are stationary, to strong collective bias where the dynamics are nonstationary and display aging up to finite-size effects. We explicitly discuss ecological implications of our work, illustrating how continuous-time renewal dynamics shape macroscopic structure and collective organization with long-term memory.
\end{abstract}

\maketitle

\section{Introduction}

Many systems exhibit long-term memory, manifested by slow, history-dependent relaxation of macroscopic observables. Examples include intermittently emitting quantum dots, neuronal activity with bursty temporal dynamics, and human or animal movement~\cite{brokmann2003statistical, bianco2007brain, gonzalez2008understanding, zaburdaev2015levy, jeon2011vivo, weigel2011ergodic, barkai2020packets, vilk2022ergodicity, vilk2022unravelling, vilk2022classification}. A predominant theoretical approach to describe such \textit{aging} behavior is based on non-Poissonian renewal processes, where broadly distributed waiting times lead to a decaying renewal rate, and observables depend explicitly on the time since system preparation~\cite{godreche2001statistics, schulz2014aging, cakir2006dynamical}. A paradigmatic realization is provided by the continuous-time random walk (CTRW), in which stochastic motion is interspersed by power-law waiting times, giving rise to anomalous transport and aging~\cite{metzler2000random, bel2005weak, metzler2014anomalous}. By contrast, Poisson renewal processes provide a memoryless benchmark: repeated renewals that reset the internal clock, restore time-translation invariance and suppress aging.

In this work we establish two fundamental features of Poisson renewal processes acting on continuous-time dynamics. First, we show that renewal dynamics formulated in continuous time can naturally give rise to a mixed discrete–continuous spatial structure. Second, we demonstrate that in an ensemble of particles under renewal, collective effects can induce persistent nonstationary behavior even under Poissonian renewal.
We focus on stochastic resetting~\cite{EM2011}, a simple realization of renewal
dynamics in which the evolution of a system is intermittently interrupted by events
that return it to a prescribed configuration. Resetting has been shown to strongly
affect transport, relaxation, and first-passage properties, and has been explored in
a wide range of contexts, including stochastic search and foraging, reaction and
enzyme kinetics, and population dynamics
\cite{reuveni2014role, kundu2024preface, koenig2021adaptive, picardi2020analysis,
blumer2022stochastic, Evansreview, gupta2022stochastic, meir2025first,
church2025accelerating}. For a single particle, CTRW under resetting to the initial position has been studied primarily in the diffusion limit \cite{kusmierz2019subdiffusive, maso2019transport}. In this regime, reinitialization of the continuous-time clock leads to dynamics that closely resemble Brownian motion with stochastic resetting. Below we show that this picture does not extend to a
general continuous-time formulation of renewal resetting even for a single particle, and that collective
interactions can fundamentally modify the impact of renewal. In particular, in contrast to the single-particle case, macroscopic memory can emerge even when all microscopic reset events are memoryless.

Resetting in many-body systems has been studied in a variety of settings, generating rich collective behavior~\cite{NagarGupta, SSM, vatash2025many, biroli2023extreme, biroli2025resetting, vilk2025collective, VAM2022, Berestycki1, Berestycki2}. While some of these studies employ renewal-based descriptions, a systematic treatment of interacting anomalous transport under reset remains challenging. Recently, an age-structured hydrodynamic framework has been proposed to describe the collective dynamics of ensembles of anomalously diffusing particles subject to renewal resets \cite{MV2025}. Here we expand and generalize this approach to derive a description of CTRW particles undergoing collective Poisson resets.

\begin{figure*}[t]
\centering
\includegraphics[clip,width=0.6\textwidth]{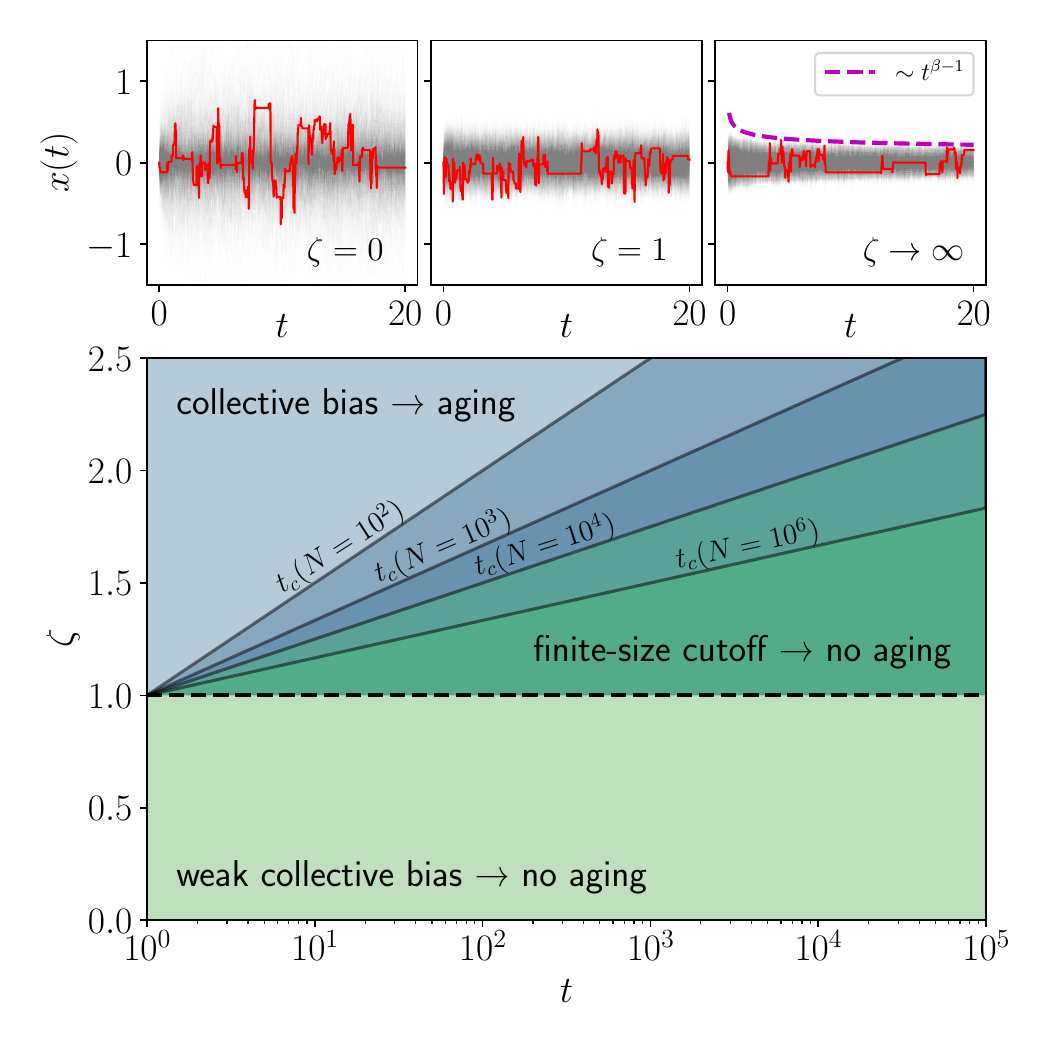}
\caption{
\textbf{Collective renewal and aging.}
Top: Simulation trajectory (1 in red overlaid on $10^3$ in gray) for reset protocols with different bias exponent $\zeta$. For $\zeta=0$, independent renewal suppresses aging; for $\zeta=1$, finite $N$ inhibits collective aging; and for strong bias $\zeta\gg1$, resets are dominated by extreme particles, leading to nonstationary system size (dashed line) and collective memory ($\beta=0.8$). Bottom: Dynamical regime diagram in the $(t,\zeta)$ plane. For $\zeta>1$, collective bias induces aging up to a finite-size crossover time $t_c\sim N^{\zeta-1}$.
}
\label{fig:schematic}
\end{figure*}

\section{Model definition and main results}
\label{sec:model}

We consider an ensemble of $N$ particles performing CTRW on the line. Each particle alternates between waiting periods and jumps, with waiting times $\tau$ sampled from a power-law distribution $\psi(\tau)$ and jump displacements $\xi$ sampled from a kernel $\omega(\xi)$. For concreteness, we take
\begin{equation}
\psi(\tau)=\beta\,\tau_0^{-1}\!\left(1+\tau/\tau_0\right)^{-1-\beta},
\quad
\omega(\xi)=\frac{1}{2\ell}e^{-|\xi|/\ell},
\label{psi_def}
\end{equation}
with $\beta>0$ and survival probability $\Psi(\tau)=(1+\tau/\tau_0)^{-\beta}$; our main results are independent of these specific choices. Reset events occur at global rate $rN$: at each reset, selected particles -- see different protocols below -- are instantaneously returned to the origin $x=0$ and their renewal clocks are reset by sampling a new waiting time $\tau$ from $\psi(\tau)$. The rate $rN$ is chosen so that in the absence of collective effects, each particle resets at rate $r$.

Our first central finding, presented in Sec.~\ref{sec:independent}, is that regardless of the reset protocol, a macroscopic fraction of particles occupies a discrete state at the reset location -- a hallmark we attribute to renewal dynamics. This discrete state is generic, does not depend on the reset protocol (see Secs. \ref{sec:extremal} and \ref{sec:collective}) or on the specific forms of $\psi(\tau)$ and $\omega(\xi)$, and vanishes only in the diffusion limit.

In Sec.~\ref{sec:extremal}, we establish that collective effects can generate long-term memory. We contrast two reset protocols: independent resets, where particles are randomly selected for reset, and extremal resets, where the particle farthest from the origin is reset. Using the age-structured framework, we show that independent resets lead to a stationary state and reduce the ensemble dynamics to those of a single particle ($N=1$). By contrast, extremal resets yield a time-dependent solution and persistent aging at long times~\footnote{Aging is defined here as the breakdown of time-translation invariance in macroscopic observables, manifested by their explicit dependence on the time since system preparation. In renewal processes, this is typically associated with a time-dependent renewal rate.}. The extremal reset rule couples otherwise independent renewal processes through a slowly evolving geometric constraint, generating collective memory.

Finally, in Sec.~\ref{sec:collective} we demonstrate that the transition to \textit{collective aging} is discontinuous.
To this end, we introduce a reset protocol with collective bias, controlled by exponent $\zeta$ that tunes the degree of collective bias toward spatially extreme particles (see definition in Sec.~\ref{sec:collective}), interpolating between independent resets at $\zeta=0$ and extremal resets at $\zeta\to\infty$, see top panels of Fig.~\ref{fig:schematic}. In the limit $N\to\infty$, the dynamics exhibit a discontinuous dynamical phase transition at $\zeta=1$, defined as a discontinuous change in the asymptotic long-time dynamics as the control parameter $\zeta$ is varied~\cite{garrahan2007dynamical, gradenigo2019first, meibohm2022finite}. The transition separates a stationary phase ($\zeta<1$) from a nonstationary aging phase ($\zeta\ge1$). At finite $N$, aging is ultimately cut off, with a crossover time $t_c\sim N^{\zeta-1}$. The resulting dynamical regimes are summarized in the bottom panel of Fig.~\ref{fig:schematic}.

\section{Independent resets}
\label{sec:independent}

For independent resets each particle is reset at a constant Poisson rate $r$, irrespective of the positions of other particles (Fig.~\ref{fig:schematic}, $\zeta=0$). For $N=1$ this model was studied in~\cite{kusmierz2019subdiffusive, maso2019transport}. We describe the collective dynamics using an age-structured theory. Denoting $n(x,\tau,t)$ as the particle density at position $x$, time $t$ and age $\tau$, we have
\begin{align}
\partial_t n(x,\tau,t)+\partial_\tau n(x,\tau,t)
= -\bigl[\beta(\tau)+r\bigr]\,n(x,\tau,t), \nonumber \\
\quad \tau>0,
\label{eq:AS_indep}
\end{align}
where 
$\beta(\tau)=\psi(\tau)/\Psi(\tau)=\beta/(\tau_0+\tau)$. 
The two loss terms describe renewals due to jumps, which reset the age \cite{berry2016quantitative}, and independent Poissonian resetting \cite{MV2025}. The boundary condition at $\tau=0$ collects both contributions,
\begin{align}
n(x,0,t)=
\int_{\mathbb{R}}\!dx'\!\int_0^\infty\!d\tau'\,
\beta(\tau')\,\omega(x-x')\,n(x',\tau',t) \nonumber \\ 
+ \, r\,\delta(x).
\label{eq:BC_indep}
\end{align}
Finally, the spatial density is 
\begin{equation} \label{eq:spatial_density1}
    C(x,t)=\int_0^\infty n(x,\tau,t)d\tau, 
\end{equation}
and satisfies mass conservation,
\begin{equation}
 \int_{\mathbb{R}}C(x,t)\,dx=1. 
\end{equation}
For $t\gg1$ we seek a steady-state $n_s(x,\tau)$. Setting $\partial_t n=0$ in
Eq.~\eqref{eq:AS_indep} reduces it to $\partial_\tau n_s = -[\beta(\tau)+r]\,n_s$, which is solved by $n_s(x,\tau)=n_s(x,0)\,\Psi(\tau)e^{-r\tau}$. Introducing the reset-modified survival integral $\tilde\mu=\int_0^\infty \Psi(\tau)e^{-r\tau}\,d\tau$, denoting $C(x)\equiv C(x,t\gg1)$, and integrating over $\tau$, gives $C(x)=n_s(x,0)\,\tilde\mu$. The steady-state solution can then be written as
\begin{equation}
n_s(x,\tau)=\frac{C(x)}{\tilde\mu}\,\Psi(\tau)\,e^{-r\tau}.
\label{eq:ns_indep}
\end{equation}
The factor $e^{-r\tau}$ in $\tilde{\mu}$ implies that resets effectively truncate the heavy-tailed waiting times, rendering $\tilde\mu$ finite for all $\beta>0$. Substituting into Eq.~\eqref{eq:BC_indep} and using $C(x)=\int_0^\infty n(x,\tau)\,d\tau$ gives a linear inhomogeneous Fredholm integral equation of the second kind,
\begin{equation}
C(x)=\hat\psi\int_{\mathbb R}\omega(x-x')\,C(x')\,dx'
+\tilde\mu\,r\,\delta(x),
\label{eq:IE_indep}
\end{equation}
where $\hat\psi=\int_0^\infty\psi(\tau)e^{-r\tau}\,d\tau$. Noting that $\omega(x)$ is the Green’s function of the operator $1-\ell^2\partial_x^2$, we apply this operator to Eq.~\eqref{eq:IE_indep} and solve for $C(x)$. This yields
\begin{align}
C(x)=[1-\hat\psi]\,\delta(x) 
+\frac{\hat\psi\sqrt{1-\hat\psi}}{2\ell}\,e^{-|x|\sqrt{1-\hat\psi}/\ell}.
\label{eq:C_indep}
\end{align}
As $\tilde\mu$ is finite, a steady state exists for any $\beta>0$. Defining $m_0$ as the delta-peak contribution to the density, Eq.~\eqref{eq:C_indep} predicts $m_0= 1-\hat\psi$. For the power-law waiting-time density in Eq.~\eqref{psi_def}, $\hat\psi$ evaluates to
\begin{equation}
\hat\psi = \beta\,e^{r\tau_0}\,E_{\beta+1}(r\tau_0),
\label{eq:hatpsi_powerlaw}
\end{equation}
where $E_\nu(z)=\int_1^\infty e^{-z u}u^{-\nu}\,du$ is the exponential integral. We note that for exponential waiting times with mean $\tau_0$,  $\hat\psi$ evaluates to
\begin{equation}
\hat\psi = \frac{1}{1+r\tau_0}.
\label{eq:hatpsi_exp}
\end{equation}
In Fig.~\ref{fig:independent_resets} we compare Eq.~\eqref{eq:C_indep} with Monte Carlo simulations (see Appendix A) for both power-law [Eq. \eqref{eq:hatpsi_powerlaw}] and exponential [Eq. \eqref{eq:hatpsi_exp}] waiting times, confirming the analytical predictions.

\begin{figure*}[t]
\centering
\includegraphics[clip,width=0.6\textwidth]{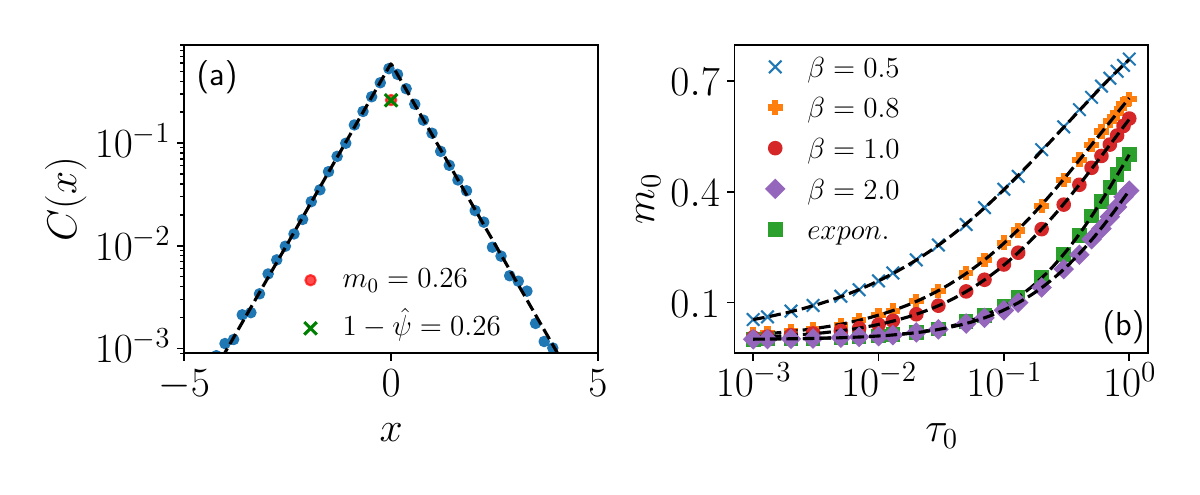}
\caption{
\textbf{Mixed discrete-continuous steady state.} (a) Simulations for the continuous density and localized mass $m_0$ (blue dots and red dot, respectively) compared to Eq.~\eqref{eq:C_indep} (dashed line and green x) for $\beta=0.8$, $\tau_0=0.1$.
(b) Localized mass $m_0$ versus $\tau_0$ for power-law [Eq.~\eqref{psi_def}] and exponential waiting times. Simulations (see legend), compared to Eq.~\eqref{eq:C_indep} (dashed line). In both panels, $\tau_0=\ell^2$, $r=1$, $N=10^4$, and $t=10^3$.    
}
\label{fig:independent_resets}
\end{figure*}

In the diffusion limit $\ell\to0$ with $D_{\mathrm{eff}}\equiv\ell^2/\tilde\mu$ fixed~\footnote{By ``diffusion limit'' we mean the continuum limit $\ell\to0$, $\tau_0\to0$ with the effective transport coefficient $D_{\mathrm{eff}}$ held fixed; this reduces the discrete CTRW to a (fractional) diffusion equation regardless of whether transport is normal or anomalous.}, the regular part of $C(x)$ reduces to
\begin{equation}
C(x)\simeq\frac{1}{2}\sqrt{\frac{r}{D_{\mathrm{eff}}}}\, e^{-|x|\sqrt{r/D_{\mathrm{eff}}}},
\label{eq:C_diffusion_limit}
\end{equation}
while the delta peak vanishes. This coincides with the result for a single CTRW particle under renewal resets \cite{kusmierz2019subdiffusive, maso2019transport}.

Our first main result is the emergence of a macroscopic delta peak in Eq.~\eqref{eq:C_indep}, corresponding to particles that have been reset but have not yet jumped. This reflects the generic coexistence of discrete and continuous components in renewal dynamics. Delta-peak contributions are known in aging renewal processes \cite{schulz2014aging}, in non-renewal resets \cite{kusmierz2019subdiffusive}, and in resets with a refractory period \cite{evans2018effects}. Here we show that it is an intrinsic consequence of the continuous-time structure and arises independently of the waiting-time law or the resetting protocol (see Secs.~\ref{sec:extremal} and~\ref{sec:collective} below), disappearing only in the diffusion limit (Fig. \ref{fig:independent_resets}b).

\section{Extremal resets}
\label{sec:extremal}

We next consider extremal resets, where at each reset event the particle farthest from the origin is moved to $x=0$ (Fig.~\ref{fig:schematic}, $\zeta\to\infty)$. This protocol was studied for Brownian and scaled Brownian particles \cite{VAM2022, MV2025}. The age-structured equation now reads,
\begin{align}
\partial_t n(x,\tau,t)+\partial_\tau n(x,\tau,t)
= -\beta(\tau)\,n(x,\tau,t),
\label{HD_bulk}
\end{align}
for $|x|<L(t)$, with $n(x, \tau, t)=0$ for $|x|>L(t)$.
Unlike independent resets, there is no bulk reset term as resets act only through the boundary. The boundary condition at $\tau=0$ collects contributions from jumps and resets within the compact support $[-L(t), L(t)]$,
\begin{align}
n(x,0,t)=\int_{-L(t)}^{L(t)}dx'\int_0^\infty d\tau'\,
\beta(\tau')\,&\omega(x-x')\,n(x',\tau',t) \nonumber  \\ 
&+ r\,\delta(x).
\label{BC_tau0}
\end{align}
The spatial density is defined as in Eq.~\eqref{eq:spatial_density1},
and satisfies mass conservation,
\begin{equation} \label{eq:normalization_extremal1}
\int_{-L(t)}^{L(t)} C(x,t)\,dx = 1.
\end{equation}

\subsection{Steady state}
\label{subsec:extremal_steady}

For $\beta>1$, the mean waiting time $\mu=\int_0^\infty\Psi(\tau)\,d\tau=\tau_0/(\beta-1)$ is finite and a stationary solution exists.
Setting $\partial_t n_s=0$ in Eq.~\eqref{HD_bulk} gives $\partial_\tau n_s=-\beta(\tau)\,n_s$, whose solution is
\begin{equation}
n_s(x,\tau)=C(x)\,\frac{\Psi(\tau)}{\mu}.
\label{eq:ns_solution}
\end{equation}
Substituting Eq.~\eqref{eq:ns_solution} into the boundary condition~\eqref{BC_tau0} yields a closed integral equation for the stationary spatial density,
\begin{align}
C(x)=\frac{1}{2\ell}\int_{-L}^{L} e^{-|x-x'|/\ell}\,C(x')\,dx'
&+ \mu r\,\delta(x),
\nonumber\\
&\qquad |x|<L,
\label{eq:IE}
\end{align}
with $\int_{-L}^{L}C(x)\,dx=1$.
We note that the boundary conditions at $|x|=L$ are \emph{not} $C(\pm L)=0$; the stationary density may be discontinuous at the edges of the support.
This contrasts with the Brownian~\cite{VAM2022} and scaled-Brownian~\cite{MV2025} cases, which are only well defined in the diffusion limit where the density is by construction continuous.
Instead, the correct boundary conditions follow from the structure of the integral operator. Defining $u(x)=\int_{-L}^{L}\omega(x-x')\,f(x')\,dx'$ for any integrable $f$, for $x\to L^-$ one has $|x-x'|=x-x'$ for all $x'\in[-L,L]$, giving $u(x)=\frac{e^{-x/\ell}}{2\ell}\int_{-L}^{L}e^{x'/\ell}f(x')\,dx'$, and hence $u'(L^-)=-(1/\ell)\,u(L^-)$. By symmetry, $u'(-L^+)=+(1/\ell)\,u(-L^+)$.

\begin{figure*}[t]
\centering
\includegraphics[clip,width=0.6\textwidth]{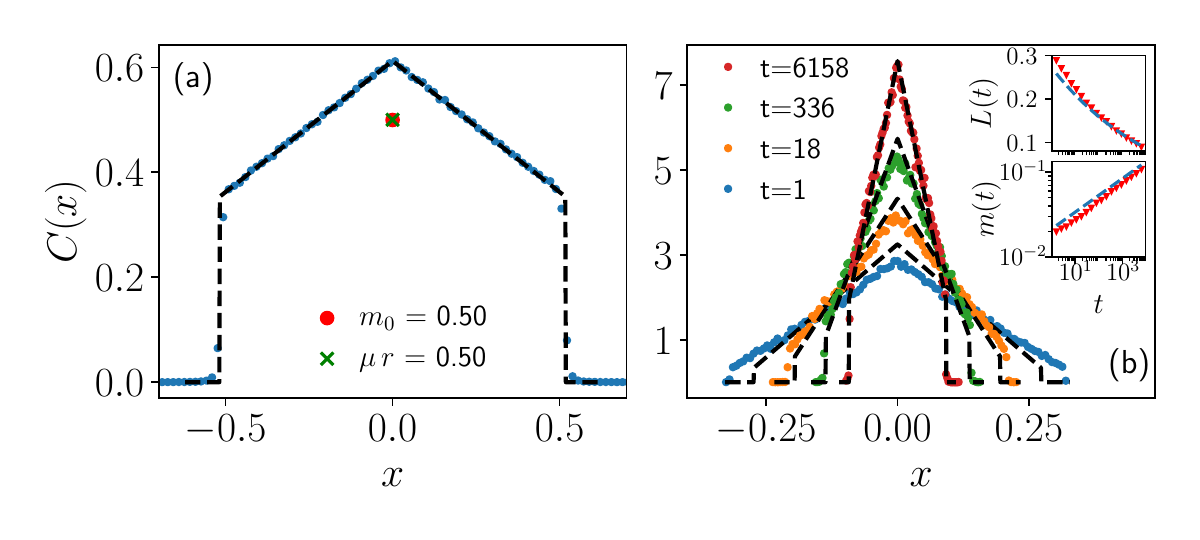}
\caption{\textbf{Extremal resets}. (a) Steady state. Simulations for the continuous density and localized mass (blue dots and red dot, respectively) compared to Eq.~\eqref{eq:Csol} (dashed line, green x) for $\beta=2$ and $\tau_0=\ell^2=0.5$. 
(b) Nonstationary regime. Simulations at different $t$ (legend) compared to Eq.~\eqref{C_adiabatic_result} (dashed lines). Insets: system size $L(t)$ and localized mass $m(t)$, Eq.~\eqref{L_beta_less_1} (dashed lines) compared to simulations (triangles), for $\beta=0.8$ and $\tau_0=\ell^2=0.001$. In both panels $N=10^5$ and $r=1$.} 
\label{fig:extremal_resetting}
\end{figure*}

To solve Eq.~\eqref{eq:IE}, we decompose $C(x)=c(x)+\mu r\,\delta(x)$, where $c(x)$ is regular on $(-L,L)$.
Noting that $\omega(x)=(2\ell)^{-1}e^{-|x|/\ell}$ is the Green's function of $1-\ell^2\partial_x^2$, we apply this operator to the equation for $c(x)$ and obtain $-\ell^2 c''(x)=\mu r\,\delta(x)$.
Hence $c$ is piecewise linear, and by symmetry $c(x)=a-b|x|$ with slope $b=\mu r/(2\ell^2)$ determined by the slope jump at $x=0$.
The intercept $a$ is fixed by evaluating $u'(L^-)=-(1/\ell)\,u(L^-)$ with $u=c-\mu r\,\omega$ and using $\omega'(L)=-(1/\ell)\,\omega(L)$, yielding $a=b(L+\ell)$.
The full stationary density is therefore
\begin{align}
C(x)=\frac{\mu r}{2\ell^2}\,(\ell+L-|x|)
+ &\mu r\,\delta(x),
\nonumber\\
&\qquad |x|<L,
\label{eq:Csol}
\end{align}
a compact triangular profile with a macroscopic delta peak $m_0 = \mu r$ at $x=0$ and a discontinuity at the boundaries $\pm L$. The support $L$ follows from mass conservation \eqref{eq:normalization_extremal1},
\begin{equation}
L=\ell\left(-1+\sqrt{\frac{2}{\mu r}-1}\right)
\simeq \sqrt{\frac{2\ell^2}{\mu r}},
\label{eq:Lsol}
\end{equation}
where the exact expression requires $\mu r\le 1$ to ensure $L\ge0$, and the final approximation holds for $\mu r\ll1$.
Comparison with simulations is shown in Fig.~\ref{fig:extremal_resetting}(a).
In the diffusion limit $\ell\ll L$ with $D_{\mathrm{eff}}=\ell^2/\mu$ fixed, Eq.~\eqref{eq:Csol} reduces to
\begin{align}
C(x)\simeq \frac{r}{2D_{\mathrm{eff}}}\,(L-|x|), \qquad
L\simeq \sqrt{\frac{2D_{\mathrm{eff}}}{r}},
\label{eq:Csol_diffusion}
\end{align}
in agreement with~\cite{VAM2022}.

\subsection{Time-dependent solution and aging}
\label{subsec:extremal_aging}

For $\beta<1$ and extremal resets, the mean waiting time diverges and no steady state exists. Equation~\eqref{HD_bulk} can be solved along characteristics for $0<\tau<t$,
\begin{equation}
n(x,\tau,t)=n(x,0,t-\tau)\,\Psi(\tau).
\label{eq:char_sol}
\end{equation}
We define the local renewal (jump) flux
\begin{equation}
J(x,t)=\int_0^\infty \beta(\tau)\,n(x,\tau,t)\,d\tau.
\label{eq:J_def}
\end{equation}
Integrating Eq.~\eqref{HD_bulk} over $\tau$ and using the boundary condition~\eqref{BC_tau0} yields the exact evolution equation
\begin{align}
\partial_t C(x,t) &= \int_{-L(t)}^{L(t)} \omega(x-x')\,J(x',t)\,dx' \nonumber\\
&\quad\quad\quad\quad\quad\quad - J(x,t) + r\,\delta(x),
\label{C_exact}
\end{align}
valid for $|x|<L(t)$.
Substituting Eq.~\eqref{eq:char_sol} into the definitions~\eqref{eq:spatial_density1} and~\eqref{eq:J_def}, and neglecting decaying initial-condition terms, gives
\begin{align}
C(x,t)&=\int_0^t \Psi(\tau)\,n(x,0,t-\tau)\,d\tau, \label{eq:C_volterra}\\
J(x,t)&=\int_0^t \psi(\tau)\,n(x,0,t-\tau)\,d\tau. \label{eq:J_volterra}
\end{align}
Laplace-transforming both equations and eliminating $\hat n(x,0,s)$ yields a closed relation between the flux and the density,
\begin{equation}
J(x,t)=\int_0^t \Phi(t-\tau)\,C(x,\tau)\,d\tau,
\label{eq:J_Phi_C}
\end{equation}
where the memory kernel $\Phi(t)$ is defined via its Laplace transform,
\begin{equation}
\hat\Phi(s)=\frac{s\,\hat\psi(s)}{1-\hat\psi(s)}.
\label{eq:Phi_Laplace}
\end{equation}
The results above are exact for all $0<\beta<1$. However, to make analytical progress, we use the small-$s$ expansion $1-\hat\psi(s)\simeq \Gamma(1-\beta)(\tau_0 s)^{\beta}$, which gives the long-time behavior
\begin{equation}
\Phi(t)\simeq -\frac{(1-\beta)\sin(\pi\beta)}{\pi\,\tau_0^{\beta}}\,
t^{\beta-2},\qquad t\to\infty.
\label{Phi_long_time}
\end{equation}
This algebraic decay reflects the aging of the underlying renewal process and induces nonstationary evolution of $C(x,t)$. 

Importantly, in the diffusion limit $\ell\to0$ with $D_{\mathrm{eff}}=\ell^{2}/\tau_0^{\beta}$ fixed, the spatial nonlocality reduces to a Laplacian, and Eq.~\eqref{C_exact} takes the form of a fractional Fokker-Planck equation on a compact support,
\begin{align}
\partial_t C(x,t)
\simeq \frac{D_{\mathrm{eff}}}{\Gamma(1-\beta)}\,D_t^{1-\beta}\partial_x^2 C(x,t)
&+r\,\delta(x),
\nonumber\\
& |x|<L(t),
\label{eq:fractional_FP}
\end{align}
with $C(\pm L(t))=0$, and where $D_t^{1-\beta}$ denotes the Caputo fractional derivative. Equation~\eqref{eq:fractional_FP} is consistent with similar long-time limits of continuous-time random walks~\cite{metzler2000random,meerschaert2004limit}.

Returning to Eq.~\eqref{eq:J_Phi_C}, for $t\gg1$ we make the a priori assumption that the density $C(x,t)$ varies slowly over the memory window of $\Phi$. Using the long-time form~\eqref{Phi_long_time}, this yields the adiabatic closure
\begin{align}
J(x,t)\simeq C(x,t)\int_0^t \Phi(u)\,du
\equiv C(x,t)\,\lambda(t),
\label{eq:adiabatic_closure}
\end{align}
with renewal rate
\begin{equation}
\lambda(t)\simeq
\frac{\sin(\pi\beta)}{\pi\tau_0^{\beta}}\,
t^{\beta-1}.
\label{lambda_asymptotic}
\end{equation}
Substituting Eq.~\eqref{eq:adiabatic_closure} into Eq.~\eqref{C_exact} and seeking a quasi-static solution with $\partial_t C\simeq 0$ yields an approximate equation identical in form to the stationary equation~\eqref{eq:IE}, under the mapping $\mu r\to r/\lambda(t)$.
The time-dependent density is therefore
\begin{align}
C(x,t)\simeq\frac{r}{2\ell^{2}\lambda(t)}\bigl(\ell+L(t)-|x|\bigr) &  +\frac{r}{\lambda(t)}\,\delta(x),
\nonumber\\
&
\qquad |x|<L(t).
\label{C_adiabatic_result}
\end{align}
Imposing mass conservation [Eq.~\eqref{eq:normalization_extremal1}] yields
\begin{equation}
L(t)=\ell\left(-1+\sqrt{\frac{2\lambda(t)}{r}-1}\right)
\simeq \ell\,\sqrt{\frac{2\lambda(t)}{r}},
\label{eq:L_adiabatic}
\end{equation}
which holds for $\lambda(t)\gg r$.
Using Eq.~\eqref{lambda_asymptotic}, we obtain the second main result of this work: the system size $L(t)$ decays and the localized mass at the delta peak $m(t)$ grows algebraically,
\begin{align}
L(t)\propto t^{(\beta-1)/2}, \qquad
m(t)= \frac{r}{\lambda(t)}\propto t^{1-\beta}.
\label{L_beta_less_1}
\end{align}
The absence of a steady state and the algebraic growth of $m(t)$ are hallmarks of aging in the underlying renewal process. A natural way to quantify this is through the aging exponent
\begin{equation}
\gamma \equiv \frac{d \log m(t)}{d\log t} = 1-\beta,
\label{eq:gamma_def}
\end{equation}
which vanishes in the stationary regime ($\beta>1$) and increases monotonically toward $1$ as $\beta\to0$.
In Fig.~\ref{fig:extremal_resetting}(b) we compare Eqs.~(\ref{C_adiabatic_result}--\ref{L_beta_less_1}) to simulations.

Two remarks are in order. First, the quasi-static solution requires $L(t)\ge0$, i.e., $\lambda(t)\ge r$. Since $\lambda(t)\to0$ for $0<\beta<1$, the support collapses at a finite time $t_\ast$ defined by $\lambda(t_\ast)=r$. Using Eq.~\eqref{lambda_asymptotic},
\begin{align}
t_\ast &\sim \left(\frac{\sin(\pi\beta)}{\pi r\,\tau_0^\beta}\right)^{1/(1-\beta)}
\propto \tau_0^{-\beta/(1-\beta)}.
\label{eq:t_collapse}
\end{align}
The adiabatic approximation holds for $t\ll t_\ast$; notably, in the diffusion limit $\tau_0\to0$ one has $t_\ast\to\infty$ and there is no finite-time collapse.

Second, the consistency of the adiabatic closure~\eqref{eq:adiabatic_closure} can be verified a posteriori. The quasi-static solution~\eqref{C_adiabatic_result} scales as $C(x,t)\sim \lambda(t)^{-1}\sim t^{1-\beta}$, so that $\partial_t C\sim t^{-\beta}\to 0$ as $t\to\infty$, confirming the quasi-static assumption. Moreover, the memory kernel~\eqref{Phi_long_time} behaves as $\Phi(t-\tau)\sim (t-\tau)^{\beta-2}$, which is sharply peaked near $\tau\approx t$; since $C(x,\tau)$ varies only weakly over the width of this peak, the factorization $\int_0^t \Phi(t-\tau)\,C(x,\tau)\,d\tau \simeq \lambda(t)\,C(x,t)$ in Eq.~\eqref{eq:adiabatic_closure} is justified.

\section{Collective bias resets}
\label{sec:collective}

To investigate the onset of collective aging, we introduce a family of collective reset protocols based on particle rank, which interpolate between independent and extremal resets. At each reset, particles are ranked by their distance from the origin, with rank $k=1$ ($k=N$) denoting the particle with the largest (smallest) $|x|$, and the probability to reset a particle decreases as $k^{-\zeta}$, so that particles farther from the origin are preferentially reset. For $\zeta=0$, all ranks are equally likely and resets are independent, while increasing $\zeta$ progressively concentrates reset events onto extreme particles, with $\zeta\to\infty$ recovering extremal resets.

In the limit $N\gg 1$, rank information is encoded through the tail mass $T(s,t)=\int_{|y|>s}C(y,t)\,dy$. The rank-based selection rule defines a weight,
\begin{align}
w(x,t)&=\bigl(1+N\,T(|x|,t)\bigr)^{-\zeta},
\label{eq:weight_rank}
\end{align}
and corresponding effective reset rate,
\begin{align}
\alpha(x,t)&=r\,\frac{w(x,t)}{Z(t)},
\label{eq:alpha_rank}
\end{align}
where $Z(t)=\int_{\mathbb R} w(x,t)\,C(x,t)\,dx$. The age-structured equation reads
\begin{equation}
\partial_t n(x,\tau,t)+\partial_\tau n(x,\tau,t)
= -\bigl[\beta(\tau)+\alpha(x,t)\bigr]\,n(x,\tau,t),
\label{eq:rank_age_pde}
\end{equation}
with boundary condition~\eqref{eq:BC_indep}. For $\zeta=0$ one recovers $\alpha(x,t)=r$ (independent resets), while for $\zeta\to\infty$ the bulk reset rate vanishes and the extremal protocol is recovered.
Despite the explicit appearance of $N$ in the microscopic rule, the factors of $N$ cancel in the ratio $\alpha=rw/Z$ at any fixed bulk position where $T(|x|,t)=O(1)$; finite-$N$ effects enter only through the extreme tail.

\begin{figure}[t]
\centering
\includegraphics[clip,width=0.4\textwidth]{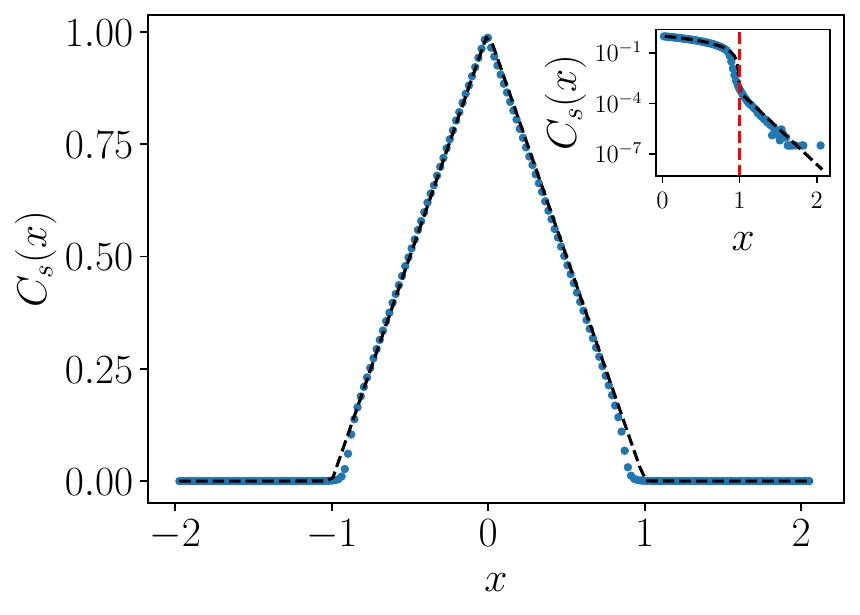}
\caption{\textbf{Rank-based resets: stationary regime.}
Stationary spatial density $C_s(x)$ for $\beta=1.5$, $\zeta=2$, from $5\times10^{4}$ independent realizations with $N=10^{4}$ particles (symbols).
Solid lines: theoretical bulk profile.
Parameters: $\tau_0=\ell^2=0.01$, $r=1$.
Inset: semi-log plot showing exponential tail decay and the predicted system size $L_N$ (dashed red line).
}
\label{fig:rank_based_stationary}
\end{figure}

\subsection{Steady state for $\beta>1$}

At stationarity, Eq.~\eqref{eq:rank_age_pde} reduces to $\partial_\tau n_s = -[\beta(\tau)+\alpha(x)]\,n_s$, which generalizes the steady-state equation of Sec.~\ref{subsec:extremal_steady} to include a position-dependent reset rate. The solution is
\begin{equation} \label{eq:steady_rank_solution_1}
    n_s(x,\tau)=n_s(x,0)\,\Psi(\tau)\,e^{-\alpha(x)\tau}.
\end{equation}
Integrating $n_s(x,\tau)$ over $\tau$ gives the spatial density $C_s(x) = n_s(x,0)\,\tilde\mu(\alpha(x))$ and similarly the stationary jump flux $J_s(x) = \hat\psi(\alpha(x))\,C_s(x)/\tilde\mu(\alpha(x))$, where we have defined 
\begin{align}
\tilde\mu(\alpha)&=\int_0^\infty \Psi(\tau)e^{-\alpha\tau}\,d\tau,
\nonumber\\
\hat\psi(\alpha)&=\int_0^\infty \psi(\tau)e^{-\alpha\tau}\,d\tau,
\end{align}
Substituting $C_s(x)$ and $J_s(x)$ into the boundary condition~\eqref{eq:BC_indep} (see also Sec.~\ref{subsec:extremal_steady}) yields the nonlinear integral equation
\begin{equation}
\frac{C_s(x)}{\tilde\mu(\alpha(x))}
=
\int_{\mathbb R}\omega(x-x')
\frac{\hat\psi(\alpha(x'))}{\tilde\mu(\alpha(x'))}
\,C_s(x')\,dx'
+
r\,\delta(x).
\label{eq:steady_rank}
\end{equation}
The nonlinearity of Eq.~\eqref{eq:steady_rank}, which arises from the self-consistent dependence of $\alpha(x)$ on $C_s$ through the tail mass $T(|x|)$, precludes a closed-form solution; while a detailed study of this equation is an interesting problem in its own right, here we focus on the far-tail and bulk limits, which are analytically tractable and capture the key physics.

\begin{figure*}[t]
\centering
\includegraphics[clip,width=0.6\textwidth]{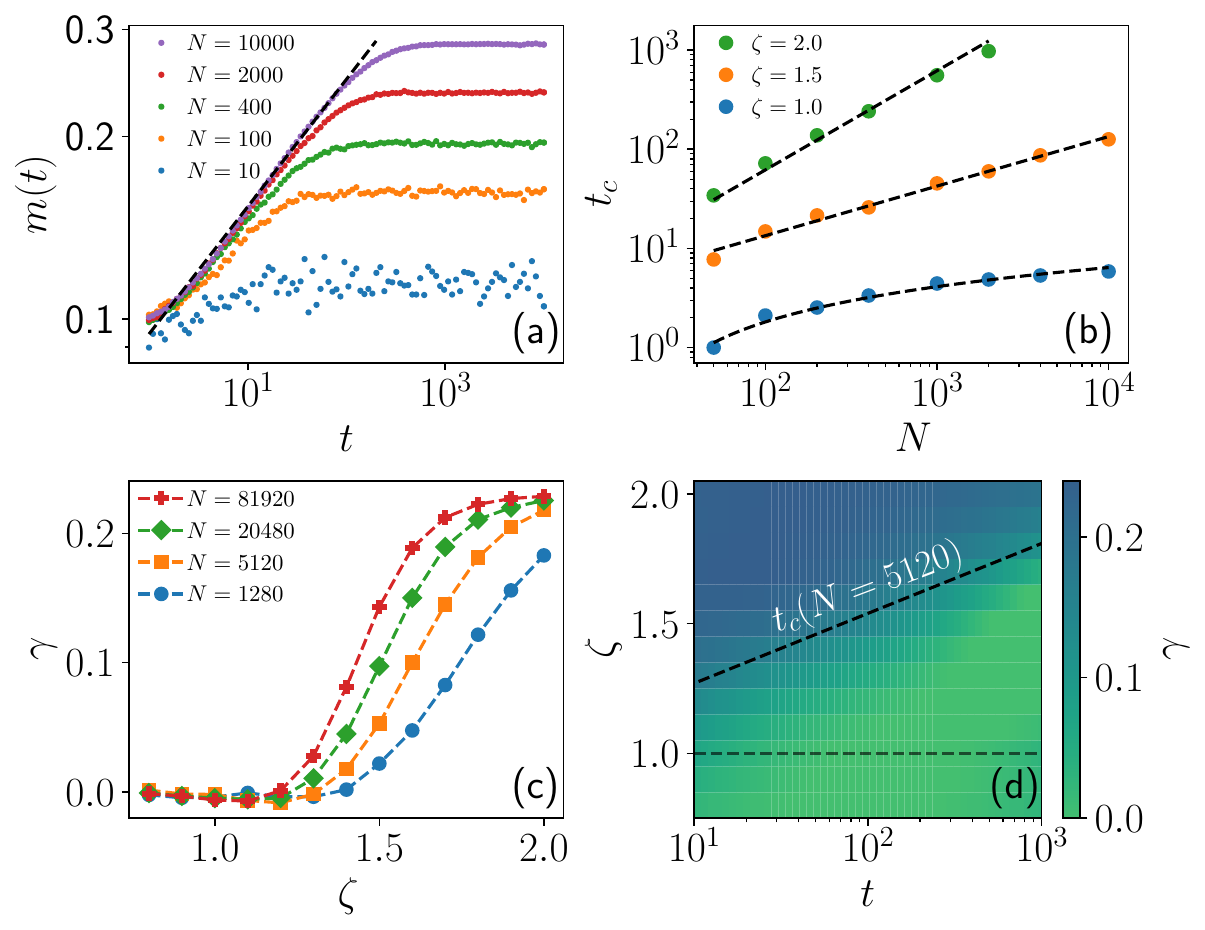}
\caption{\textbf{Dynamical phase transition induced by collective bias.}
(a) Localized mass $m(t)$ for different $N$ (see legend) and $\zeta=1.5$, showing an aging regime followed by saturation due to finite-size effects. Compared to the theoretical prediction $m(t)\sim t^{1-\beta}$ (dashed line). (b) Finite-size cutoff time $t_c$ versus $N$ for different $\zeta$ (see legend), with the scalings $t_c\sim\log N$ at $\zeta=1$ and $t_c\sim N^{\zeta-1}$ for $\zeta>1$ (dashed lines).
 (c) Aging exponent $\gamma\equiv d \log m(t)/d\log t$ as a function of $\zeta$, for different $N$ at $t=10^2$. (d) Map of $\gamma$ versus $\zeta$ and $t$ (see colorbar) for $N=5120$. Dashed lines denote $\zeta =1$ and the theoretical $t_c \sim N^{\zeta -1}$. In all panels, $\beta=0.8$, $\tau_0 = \ell^2 = 0.01$, and $r=1$. } 
\label{fig:rank_based_resets}
\end{figure*}

In the far tail, $T(|x|)\to0$ and $\alpha(x)\to \alpha_\infty=r/Z$. Using the substitution $u=T(s)$ (so that $du=-2C_s(s)\,ds$, which cancels the density in the integrand of $Z$), the normalization evaluates to $Z=\int_0^1(1+Nu)^{-\zeta}\,du$, giving for large $N$,
\begin{equation}
\alpha_\infty=\frac{r}{Z}\sim
\begin{cases}
r(1-\zeta)\,N^{\zeta}, & \zeta<1,\\[4pt]
rN/\log N, & \zeta=1,\\[4pt]
r(\zeta-1)\,N, & \zeta>1.
\end{cases}
\label{eq:alpha_inf}
\end{equation}
Thus, for $T(|x|)\to0$ Eq.~\eqref{eq:steady_rank} becomes linear with constant $\alpha_\infty$. Applying the operator $(1-\ell^2\partial_x^2)$ yields $\ell^2 g''(x) = (1-\hat\psi(\alpha_\infty))\,g(x)$ with $g = C_s/\tilde\mu(\alpha_\infty)$, whose solution gives
\begin{align}
C_s(x)\propto e^{-\kappa_\infty|x|},
\qquad
\kappa_\infty=\frac{1}{\ell}\sqrt{1-\hat\psi(\alpha_\infty)}.
\label{eq:tail_decay}
\end{align}

In the bulk, i.e.\ at fixed $|x|=O(1)$ as $N\to\infty$, $T(|x|)=O(1)$ and $w(x)\sim N^{-\zeta}\,T(|x|)^{-\zeta}$, so that the bulk reset rate is
\begin{equation}
\alpha_{\mathrm{bulk}}=\frac{r\,w}{Z}\sim
\begin{cases}
r(1-\zeta)\,T(|x|)^{-\zeta}, & \zeta<1,\\[4pt]
r\,T(|x|)^{-1}/\log N, & \zeta=1,\\[4pt]
r(\zeta-1)\,N^{1-\zeta}\,T(|x|)^{-\zeta}, & \zeta>1.
\end{cases}
\label{eq:alpha_bulk}
\end{equation}
Defining $g(x)=C_s(x)/\tilde\mu(\alpha(x))$ and applying the operator $(1-\ell^2\partial_x^2)$ to Eq.~\eqref{eq:steady_rank} yields the exact local equation
\begin{equation}
\ell^2 g''(x)=[1-\hat\psi(\alpha(x))]\,g(x), \qquad x\neq0,
\label{eq:g_exact}
\end{equation}
where the nonlinearity is entirely contained in the self-consistent dependence $\hat\psi(\alpha(x))$ with $\alpha(x)=r\,w(x)/Z$.
Equation~\eqref{eq:alpha_bulk} reveals two qualitatively different regimes.

\textit{Regime $\zeta>1$: extremal-like bulk.} For $\zeta>1$, $\alpha_{\mathrm{bulk}}\to0$ as $N\to\infty$, so renewal resetting becomes asymptotically ineffective in the bulk while remaining strong in the far tail.
Since $\hat\psi(\alpha)=1-\mu\alpha+o(\alpha)$ and $\tilde\mu(\alpha)\to\mu$ for small $\alpha$, in the bulk $\hat\psi\to1$ and Eq.~\eqref{eq:g_exact} reduces to $g''(x)\simeq 0$, giving a piecewise-linear profile $g(x)\simeq a-b|x|$.
The slope $b$ is fixed by the injection at the origin: including the $r\delta(x)$ source gives the jump condition $g'(0^+)-g'(0^-)=-r/\ell^2$, hence $b=r/(2\ell^2)$.
The linear core extends up to $|x|\simeq L_N$, beyond which the solution crosses over to the exponential tail.
Matching $g$ and $g'$ at $|x|=L_N$ by continuity yields $a\simeq b(L_N+\kappa_\infty^{-1})$, and mass conservation gives
\begin{equation}
L_N \simeq -\frac{1}{\kappa_\infty}
+ \sqrt{\frac{1}{\kappa_\infty^2}+\frac{2\ell^2}{\mu r}}.
\label{eq:LN}
\end{equation}
For $L_N\gg\kappa_\infty^{-1}$ this reduces to $L_N\simeq \sqrt{2\ell^2/(\mu r)}$, recovering the extremal-resetting scaling~\eqref{eq:Lsol}. 
In Fig.~\ref{fig:rank_based_stationary} we compare the theoretical predictions with simulations for $\beta=1.5$ and $\zeta=2$, showing good agreement for the bulk profile and the onset of the exponential tail.

\textit{Regime $\zeta<1$: strong resetting throughout.} For $\zeta<1$, the effective reset rate remains $O(1)$ throughout the bulk, so the piecewise-linear reduction no longer applies and the stationary profile is controlled by a smooth, spatially varying balance between transport and resetting. A natural starting point for this regime is a WKB-type approximation, in which the stationary profile is written as $C_s(x)\sim \tilde\mu(\alpha(x))\exp[-S(x)]$ with $S'(x)$ determined locally by the effective reset rate; a full self-consistent analysis is left for future work.

\subsection{Dynamical phase transition for $\beta<1$}

As in Eq.~\eqref{eq:steady_rank_solution_1}, solving Eq.~\eqref{eq:rank_age_pde} along characteristics shows that $\alpha(x,t)$ induces an effective exponential cutoff of the waiting-time distribution,
\begin{equation}
\Psi_{\mathrm{eff}}(\tau|x,t)=\Psi(\tau)\,e^{-\alpha(x,t)\tau},
\label{eq:Psi_eff}
\end{equation}
so that the asymptotic value of $\alpha$ in the bulk controls whether long waiting times are truncated. Equation~\eqref{eq:alpha_bulk} shows that $\alpha_{\mathrm{bulk}}\to0$ for all $\zeta\ge1$ as $N\to\infty$, so that reset events become asymptotically ineffective in the bulk while remaining strong in the tail, where extreme particles dominate the reset statistics. Focusing on the regime $\beta <1$, this underlies a discontinuous dynamical phase transition at $\zeta=1$: vanishing $\alpha_{\mathrm{bulk}}$ implies that resets fail to truncate long waiting times in the bulk, the renewal rate decays algebraically, and the dynamics become nonstationary and exhibit aging. Consequently, the aging exponent $\gamma$ jumps discontinuously from $0$ to $1-\beta$ at $\zeta=1$.

For any finite $N$, however, $\alpha_{\mathrm{bulk}}$ remains nonzero, and Eq.~\eqref{eq:Psi_eff} implies a finite cutoff time
\begin{equation}
t_c\sim\frac{1}{\alpha_{\mathrm{bulk}}}
\sim
\begin{cases}
\log N, & \zeta=1,\\[4pt]
N^{\zeta-1}, & \zeta>1,
\end{cases}
\label{eq:t_c}
\end{equation}
beyond which stationarity is eventually restored.
Figure~\ref{fig:rank_based_resets} summarizes these results for $\beta=0.8$, illustrating aging of the localized mass $m(t)$ for $1\ll t< t_c$ [Fig.~\ref{fig:rank_based_resets}(a)] and the scaling of the finite-size cutoff time $t_c$ with $N$ [Fig.~\ref{fig:rank_based_resets}(b)]. We further show that at fixed time, the dynamical transition sharpens and approaches $\zeta=1$ as $N$ increases [Fig.~\ref{fig:rank_based_resets}(c)]. Finally, for $\zeta<1$ or $t>t_c$ the dynamics are stationary with $\gamma\simeq0$, while for $\zeta>1$ and $t<t_c$ we have a sharp transition to aging with $\gamma\simeq1-\beta$ [Fig.~\ref{fig:rank_based_resets}(d)].

\section{Discussion}
\label{sec:discussion}

We have studied ensembles of CTRW particles subject to collective Poisson renewal resets, deriving exact descriptions for three resetting protocols. For independent resets, a stationary state always exists, with a mixed discrete--continuous spatial density. This mixed state is found for all reset protocols, and is absent only in the diffusion limit. For extremal resets, the collective constraint retains memory despite memoryless renewal: when $\beta<1$, the renewal rate decays algebraically and the support collapses in finite time. For rank-based resets, a discontinuous dynamical phase transition emerges at $\zeta=1$, separating stationary from aging dynamics, with a finite-size cutoff time that diverges algebraically with $N$.

As an illustrative ecological example, we note that CTRW provide a consistent description of breeding white stork movement~\cite{vilk2022classification, vilk2022unravelling}. Storks perform $\sim$3 daily foraging bouts with patch residence times of $30$--$70$ min and inter-site distances of $200$--$400$ m~\cite{alonso1991habitat}, corresponding to $\tau_0\in[0.5,1.2]$ h, $\ell\in[0.2,0.4]$ km, and $r\sim3~\mathrm{day}^{-1}$. Importantly, these values place the system far from the diffusion limit: using the estimates above, $r\tau_0\sim0.06$--$0.15$, so the localized mass $m_0=1-\hat\psi$ is appreciable (cf.\ Fig.~\ref{fig:independent_resets}b). The predicted mixed discrete--continuous spatial structure is thus directly relevant, with the delta-peak component corresponding to birds at the nest and the continuous density describing foraging excursions in the home range~\cite{zurell2018home}. 

Moreover, foraging in this species is strongly shaped by collective decision-making: storks forage in flocks and adjust patch departure times based on the behavior of conspecifics~\cite{blanco1996population, flack2018local}. This social influence introduces effective correlations between individual renewal processes. While the rank-based reset protocol is a simplified toy model that does not capture the full complexity of collective foraging decisions, it distills the essential ingredient: a tunable bias that preferentially acts on spatially extreme individuals. The rank exponent $\zeta$ may thus be interpreted as a phenomenological measure of the degree to which collective decisions concentrate on the most distant foragers. Our analysis predicts that even moderate collective bias ($\zeta\ge1$) qualitatively alters the long-time statistics, inducing aging and nonstationarity---a prediction that is robust to the specific form of the bias. More broadly, the framework suggests that long-term tracking data in collectively foraging species may exhibit signatures of history-dependent dynamics that cannot be explained by single-particle renewal models alone.

Although we focus on resetting, our results demonstrate a more general principle: interactions or global constraints in continuous-time dynamics can fundamentally alter memoryless renewal processes, allowing non-Markovian effects to emerge at the macroscopic level. These results are relevant to other renewal processes, including population dynamics with prolonged sojourns at local extinction~\cite{lande1998extinction}, neuronal spiking models with refractory renewal statistics~\cite{burkitt2006review}, and learning processes driven by restarts~\cite{kautz2002dynamic}.

\bigskip
\noindent
{\bf Acknowledgments}.  We are grateful to Baruch Meerson, Eli Barkai, and Naftali Smith for invaluable discussions. 

\bigskip
\noindent
{\bf Data availability}. The simulation code used in this work is publicly available at \url{https://gitlab.com/ohad.vilk/macroscopic_localization_and_collective_memory_in_poisson_renewal_resetting}.

\begin{appendices}

\section{Monte Carlo simulations}
\label{app:mc}

We simulate $N$ continuous-time random walkers on $\mathbb{R}$ using an event-driven algorithm \cite{gillespie1976general, gillespie1977exact}. Each particle $i$ starts at $x_i(0)=0$ and evolves via renewal waiting times and symmetric jumps:
\begin{enumerate}
\item Inter-jump times are i.i.d.\ draws from either (i) an exponential law with mean $\tau_0$, or (ii) a power-law (Pareto-type) law with survival function $\Pr(\tau>t)=(1+t/\tau_0)^{-\beta}$ (scale $\tau_0>0$, tail exponent $\beta>0$).
\item Given a jump event at time $t$, the position updates as $x_i(t^+)=x_i(t^-)+\Delta x$, where $\Delta x$ is sampled from a symmetric jump kernel with scale parameter $\ell$. We have implemented Gaussian, two-sided exponential, and discrete $\pm 1$ kernels; all figures use the two-sided exponential kernel, but we have verified that none of the main results change when using a Gaussian kernel.
\end{enumerate}
The simulation advances time by always executing the next scheduled event (a particle jump or a reset), selected by minimum time from a priority queue. After each jump, a new waiting time is drawn for that particle and its next jump is rescheduled.

Resets occur according to an independent Poisson clock. At a reset time $t_r$, one particle is selected by a rule (below) and its position is instantaneously reset to the origin:
\[
x_{i^\star}(t_r^+)=0.
\]
The reset particle's next jump time is rescheduled according to the laws above.
We implemented the three reset selection rules described in the main text:
\begin{enumerate}
  \item Random reset: choose $i^\star$ uniformly from $\{1,\dots,N\}$.
  \item Most-distant reset: choose $i^\star=\arg\max_i |x_i(t_r^-)|$.
  \item Rank-based reset: order particles by decreasing $|x_i(t_r^-)|$ to obtain ranks $k=1,\dots,N$. Draw a rank $k$ with probability $P(k)\propto k^{-\zeta}$; set $i^\star$ to that particle.
\end{enumerate}

The simulator is implemented in C++ and exposed to Python via \texttt{pybind11}. 

\end{appendices}

\bibliography{references}

@article{EM2011,
  title={Diffusion with stochastic resetting},
  author={Evans, Martin R. and Majumdar, Satya N.},
  journal={Phys. Rev. Lett.},
  volume={106},
  number={16},
  pages={160601},
  year={2011},
  publisher={APS}
}

@article{MV2025,
  title={Age-structured hydrodynamics of ensembles of anomalously diffusing particles with 
         renewal resetting},
  author={Meerson, Baruch and Vilk, Ohad},
  journal={Phys. Rev. Res.},
  volume={8},
  number={2},
  pages={023103},
  year={2026}
}

@article{vilk2022unravelling,
  title={Unravelling the origins of anomalous diffusion: From molecules to migrating storks},
  author={Vilk, Ohad and Aghion, Erez and Avgar, Tal and Beta, Carsten and Nagel, Oliver and Sabri, Adal and Sarfati, Raphael and Schwartz, Daniel K and Weiss, Matthias and Krapf, Diego and others},
  journal={Phys. Rev. Res.},
  volume={4},
  number={3},
  pages={033055},
  year={2022},
  publisher={APS}
}

@article{SSM,
  title={Fluctuations of a swarm of {B}rownian bees},
  author={Siboni, Maor and Sasorov, Pavel and Meerson, Baruch},
  journal={Phys. Rev. E},
  volume={104},
  number={5},
  pages={054131},
  year={2021},
  publisher={APS}
}

@article{Evansreview,
  title={Stochastic resetting and applications},
  author={Evans, Martin R and Majumdar, Satya N and Schehr, Gr{\'e}gory},
  journal={J. Phys. A: Math. Theor.},
  volume={53},
  number={19},
  pages={193001},
  year={2020},
  publisher={IOP Publishing}
}

@article{NagarGupta,
  title={Stochastic resetting in interacting particle systems: A review},
  author={Nagar, Apoorva and Gupta, Shamik},
  journal={J. Phys. A: Math. Theor.},
  volume={56},
  number={28},
  pages={283001},
  year={2023},
  publisher={IOP Publishing}
}

@article{kundu2024preface,
  title={Preface: stochastic resetting—theory and applications},
  author={Kundu, Anupam and Reuveni, Shlomi},
  journal={J. Phys. A: Math. Theor.},
  volume={57},
  number={6},
  pages={060301},
  year={2024},
  publisher={IOP Publishing}
}

@article{gupta2022stochastic,
  title={Stochastic resetting: A (very) brief review},
  author={Gupta, Shamik and Jayannavar, Arun M},
  journal={Front. Phys.},
  volume={10},
  pages={789097},
  year={2022},
  publisher={Frontiers Media SA}
}

@article{vatash2025many,
  title={Many-body colloidal dynamics under stochastic resetting: Competing effects of particle interactions on the steady-state distribution},
  author={Vatash, Ron and Roichman, Yael},
  journal={Phys. Rev. Res.},
  volume={7},
  number={3},
  pages={L032020},
  year={2025},
  publisher={APS}
}

@article{church2025accelerating,
  title={Accelerating molecular dynamics through informed resetting},
  author={Church, Jonathan R and Blumer, Ofir and Keidar, Tommer D and Ploutno, Leo and Reuveni, Shlomi and Hirshberg, Barak},
  journal={J. Chem. Theory Comput.},
  volume={21},
  number={2},
  pages={605--613},
  year={2025},
  publisher={ACS Publications}
}

@article{meir2025first,
  title={First-passage approach to optimizing perturbations for improved training of machine learning models},
  author={Meir, Sagi and Keidar, Tommer D and Reuveni, Shlomi and Hirshberg, Barak},
  journal={Mach. Learn.: Sci. Technol.},
  volume={6},
  number={2},
  pages={025053},
  year={2025},
  publisher={IOP Publishing}
}

@article{Berestycki2,
  title={A free boundary problem arising from branching {B}rownian motion with selection},
  author={Berestycki, Julien and Brunet, {\'E}ric and Nolen, James and Penington, Sarah},
  journal={Trans. Am. Math. Soc.},
  volume={374},
  number={09},
  pages={6269--6329},
  year={2021}
}

@article{Berestycki1,
  title={{B}rownian bees in the infinite swarm limit},
  author={Berestycki, Julien and Brunet, {\'E}ric and Nolen, James and Penington, Sarah},
  journal={Ann. Probab.},
  volume={50},
  number={6},
  pages={2133--2177},
  year={2022},
  publisher={Institute of Mathematical Statistics}
}

@article{VAM2022,
  title={Fluctuations and first-passage properties of systems of {B}rownian particles with reset},
  author={Vilk, Ohad and Assaf, Michael and Meerson, Baruch},
  journal={Phys. Rev. E},
  volume={106},
  number={2},
  pages={024117},
  year={2022},
  publisher={APS}
}

@Article{gillespie1977exact,
  author    = {Gillespie, Daniel T},
  title     = {Exact stochastic simulation of coupled chemical reactions},
  journal   = {J. Phys. Chem.},
  year      = {1977},
  volume    = {81},
  number    = {25},
  pages     = {2340--2361},
  publisher = {ACS Publications},
}

@Article{gillespie1976general,
  author    = {Gillespie, Daniel T},
  title     = {A general method for numerically simulating the stochastic time evolution of coupled chemical reactions},
  journal   = {J. Comput. Phys.},
  year      = {1976},
  volume    = {22},
  number    = {4},
  pages     = {403--434},
  publisher = {Elsevier},
}

@article{metzler2000random,
  title={The random walk's guide to anomalous diffusion: a fractional dynamics approach},
  author={Metzler, Ralf and Klafter, Joseph},
  journal={Phys. Rep.},
  volume={339},
  number={1},
  pages={1--77},
  year={2000},
  publisher={Elsevier}
}

@article{jeon2011vivo,
  title={In vivo anomalous diffusion and weak ergodicity breaking of lipid granules},
  author={Jeon, Jae-Hyung and Tejedor, Vincent and Burov, Stas and Barkai, Eli and Selhuber-Unkel, Christine and Berg-S{\o}rensen, Kirstine and Oddershede, Lene and Metzler, Ralf},
  journal={Phys. Rev. Lett.},
  volume={106},
  number={4},
  pages={048103},
  year={2011},
  publisher={APS}
}

@article{metzler2014anomalous,
  title={Anomalous diffusion models and their properties: non-stationarity, non-ergodicity, and ageing at the centenary of single particle tracking},
  author={Metzler, Ralf and Jeon, Jae-Hyung and Cherstvy, Andrey G and Barkai, Eli},
  journal={Phys. Chem. Chem. Phys.},
  volume={16},
  number={44},
  pages={24128--24164},
  year={2014},
  publisher={Royal Society of Chemistry}
}

@article{vilk2022ergodicity,
  title={Ergodicity breaking in area-restricted search of avian predators},
  author={Vilk, Ohad and Orchan, Yotam and Charter, Motti and Ganot, Nadav and Toledo, Sivan and Nathan, Ran and Assaf, Michael},
  journal={Phys. Rev. X},
  volume={12},
  number={3},
  pages={031005},
  year={2022},
  publisher={APS}
}

@article{biroli2023extreme,
  title={Extreme statistics and spacing distribution in a {B}rownian gas correlated by resetting},
  author={Biroli, Marco and Larralde, Hernan and Majumdar, Satya N and Schehr, Gr{\'e}gory},
  journal={Phys. Rev. Lett.},
  volume={130},
  number={20},
  pages={207101},
  year={2023},
  publisher={APS}
}

@article{maso2019transport,
  title={Transport properties and first-arrival statistics of random motion with stochastic reset times},
  author={Mas{\'o}-Puigdellosas, Axel and Campos, Daniel and M{\'e}ndez, Vicen{\c{c}}},
  journal={Phys. Rev. E},
  volume={99},
  number={1},
  pages={012141},
  year={2019},
  publisher={APS}
}

@article{biroli2025resetting,
  title={Resetting Dyson {B}rownian motion},
  author={Biroli, Marco and Majumdar, Satya N and Schehr, Gr{\'e}gory},
  journal={Phys. Rev. E},
  volume={112},
  number={1},
  pages={014101},
  year={2025},
  publisher={APS}
}

@article{berry2016quantitative,
  title={Quantitative convergence towards a self-similar profile in an age-structured renewal equation for subdiffusion},
  author={Berry, Hugues and Lepoutre, Thomas and Gonz{\'a}lez, {\'A}lvaro Mateos},
  journal={Acta Appl. Math.},
  volume={145},
  number={1},
  pages={15--45},
  year={2016},
  publisher={Springer}
}

@article{schulz2014aging,
  title={Aging renewal theory and application to random walks},
  author={Schulz, Johannes HP and Barkai, Eli and Metzler, Ralf},
  journal={Phys. Rev. X},
  volume={4},
  number={1},
  pages={011028},
  year={2014},
  publisher={APS}
}

@article{kusmierz2019subdiffusive,
  title={Subdiffusive continuous-time random walks with stochastic resetting},
  author={Ku{\'s}mierz, {\L}ukasz and Gudowska-Nowak, Ewa},
  journal={Phys. Rev. E},
  volume={99},
  number={5},
  pages={052116},
  year={2019},
  publisher={APS}
}

@article{meerschaert2004limit,
  title={Limit theorems for continuous-time random walks with infinite mean waiting times},
  author={Meerschaert, Mark M and Scheffler, Hans-Peter},
  journal={J. Appl. Probab.},
  volume={41},
  number={3},
  pages={623--638},
  year={2004},
  publisher={Cambridge University Press}
}

@article{godreche2001statistics,
  title={Statistics of the occupation time of renewal processes},
  author={Godr{\`e}che, C and Luck, JM},
  journal={J. Stat. Phys.},
  volume={104},
  number={3},
  pages={489--524},
  year={2001},
  publisher={Springer}
}

@article{barkai2020packets,
  title={Packets of diffusing particles exhibit universal exponential tails},
  author={Barkai, Eli and Burov, Stanislav},
  journal={Phys. Rev. Lett.},
  volume={124},
  number={6},
  pages={060603},
  year={2020},
  publisher={APS}
}

@article{picardi2020analysis,
  title={Analysis of movement recursions to detect reproductive events and estimate their fate in central place foragers},
  author={Picardi, Simona and Smith, Brian J and Boone, Matthew E and Frederick, Peter C and Cecere, Jacopo G and Rubolini, Diego and Serra, Lorenzo and Pirrello, Simone and Borkhataria, Rena R and Basille, Mathieu},
  journal={Mov. Ecol.},
  volume={8},
  number={1},
  pages={24},
  year={2020},
  publisher={Springer}
}

@inproceedings{koenig2021adaptive,
  title={Adaptive restarts for stochastic synthesis},
  author={Koenig, Jason R and Padon, Oded and Aiken, Alex},
  booktitle={Proceedings of the 42nd ACM SIGPLAN International Conference on Programming Language Design and Implementation},
  pages={696--709},
  year={2021}
}

@article{blumer2022stochastic,
  title={Stochastic resetting for enhanced sampling},
  author={Blumer, Ofir and Reuveni, Shlomi and Hirshberg, Barak},
  journal={J. Phys. Chem. Lett.},
  volume={13},
  number={48},
  pages={11230--11236},
  year={2022},
  publisher={ACS Publications}
}

@article{reuveni2014role,
  title={The role of substrate unbinding in michaelis-menten enzymatic reactions},
  author={Reuveni, Shlomi and Urbakh, Michael and Klafter, Joseph},
  journal={Biophys. J.},
  volume={106},
  number={2},
  pages={677a},
  year={2014},
  publisher={Elsevier}
}

@article{brokmann2003statistical,
  title={Statistical aging and nonergodicity in the fluorescence of single nanocrystals},
  author={Brokmann, Xavier and Hermier, J-P and Messin, Ga{\"e}tan and Desbiolles, Pierre and Bouchaud, J-P and Dahan, Maxime},
  journal={Phys. Rev. Lett.},
  volume={90},
  number={12},
  pages={120601},
  year={2003},
  publisher={APS}
}

@article{cakir2006dynamical,
  title={Dynamical origin of memory and renewal},
  author={Cakir, RASIT and Grigolini, Paolo and Krokhin, Arkadii A},
  journal={Phys. Rev. E},
  volume={74},
  number={2},
  pages={021108},
  year={2006},
  publisher={APS}
}

@article{bianco2007brain,
  title={Brain, music, and non-Poisson renewal processes},
  author={Bianco, Simone and Ignaccolo, Massimiliano and Rider, Mark S and Ross, Mary J and Winsor, Phil and Grigolini, Paolo},
  journal={Phys. Rev. E},
  volume={75},
  number={6},
  pages={061911},
  year={2007},
  publisher={APS}
}

@article{gonzalez2008understanding,
  title={Understanding individual human mobility patterns},
  author={Gonzalez, Marta C and Hidalgo, Cesar A and Barabasi, Albert-Laszlo},
  journal={Nature},
  volume={453},
  number={7196},
  pages={779--782},
  year={2008},
  publisher={Nature Publishing Group UK London}
}

@article{zaburdaev2015levy,
  title={L{\'e}vy walks},
  author={Zaburdaev, Vasily and Denisov, Sergey and Klafter, Joseph},
  journal={Rev. Mod. Phys.},
  volume={87},
  number={2},
  pages={483--530},
  year={2015},
  publisher={APS}
}

@article{weigel2011ergodic,
  title={Ergodic and nonergodic processes coexist in the plasma membrane as observed by single-molecule tracking},
  author={Weigel, Aubrey V and Simon, Blair and Tamkun, Michael M and Krapf, Diego},
  journal={Proc. Natl. Acad. Sci. U.S.A.},
  volume={108},
  number={16},
  pages={6438--6443},
  year={2011},
  publisher={National Acad Sciences}
}

@article{bel2005weak,
  title={Weak ergodicity breaking in the continuous-time random walk},
  author={Bel, Golan and Barkai, Eli},
  journal={Phys. Rev. Lett.},
  volume={94},
  number={24},
  pages={240602},
  year={2005},
  publisher={APS}
}

@article{vilk2025collective,
  title={Collective behavior of independent scaled {B}rownian particles with renewal resetting},
  author={Vilk, Ohad and Meerson, Baruch},
  journal={Physica A},
  volume={},
  pages={131542},
  year={2026}
}

@article{vilk2022classification,
  title={Classification of anomalous diffusion in animal movement data using power spectral analysis},
  author={Vilk, Ohad and Aghion, Erez and Nathan, Ran and Toledo, Sivan and Metzler, Ralf and Assaf, Michael},
  journal={J. Phys. A: Math. Theor.},
  volume={55},
  number={33},
  pages={334004},
  year={2022},
  publisher={IOP Publishing}
}

@article{evans2018effects,
  title={Effects of refractory period on stochastic resetting},
  author={Evans, Martin R and Majumdar, Satya N},
  journal={J. Phys. A: Math. Theor.},
  volume={52},
  number={1},
  pages={01LT01},
  year={2018},
  publisher={IOP Publishing}
}

@article{meibohm2022finite,
  title={Finite-time dynamical phase transition in nonequilibrium relaxation},
  author={Meibohm, Jan and Esposito, Massimiliano},
  journal={Phys. Rev. Lett.},
  volume={128},
  number={11},
  pages={110603},
  year={2022},
  publisher={APS}
}

@article{gradenigo2019first,
  title={A first-order dynamical transition in the displacement distribution of a driven run-and-tumble particle},
  author={Gradenigo, Giacomo and Majumdar, Satya N},
  journal={J. Stat. Mech. Theory Exp.},
  volume={2019},
  number={5},
  pages={053206},
  year={2019},
  publisher={IOP Publishing}
}

@article{garrahan2007dynamical,
  title={Dynamical first-order phase transition in kinetically constrained models of glasses},
  author={Garrahan, Juan P and Jack, Robert L and Lecomte, Vivien and Pitard, Estelle and van Duijvendijk, Kristina and van Wijland, Fr{\'e}d{\'e}ric},
  journal={Phys. Rev. Lett.},
  volume={98},
  number={19},
  pages={195702},
  year={2007},
  publisher={APS}
}

@article{alonso1991habitat,
  title={Habitat selection by foraging White Storks, Ciconia ciconia, during the breeding season},
  author={Alonso, Juan C and Alonso, Javier A and Carrascal, Luis M},
  journal={Can. J. Zool.},
  volume={69},
  number={7},
  pages={1957--1962},
  year={1991},
  publisher={NRC Research Press Ottawa, Canada}
}

@article{zurell2018home,
  title={Home range size and resource use of breeding and non-breeding white storks along a land use gradient},
  author={Zurell, Damaris and Von Wehrden, Henrik and Rotics, Shay and Kaatz, Michael and Gro{\ss}, Helge and Schlag, Lena and Sch{\"a}fer, Merlin and Sapir, Nir and Turjeman, Sondra and Wikelski, Martin and others},
  journal={Front. Ecol. Evol.},
  volume={6},
  pages={79},
  year={2018},
  publisher={Frontiers Media SA}
}

@article{blanco1996population,
  title={Population dynamics and communal roosting of white storks foraging at a Spanish refuse dump},
  author={Blanco, Guillermo},
  journal={Colonial Waterbirds},
  pages={273--276},
  year={1996},
  publisher={JSTOR}
}

@article{flack2018local,
  title={From local collective behavior to global migratory patterns in white storks},
  author={Flack, Andrea and Nagy, M{\'a}t{\'e} and Fiedler, Wolfgang and Couzin, Iain D and Wikelski, Martin},
  journal={Science},
  volume={360},
  number={6391},
  pages={911--914},
  year={2018},
  publisher={American Association for the Advancement of Science}
}

@article{kautz2002dynamic,
  title={Dynamic restart policies},
  author={Kautz, Henry and Horvitz, Eric and Ruan, Yongshao and Gomes, Carla and Selman, Bart},
  journal={AAAI/IAAI},
  volume={97},
  pages={674--681},
  year={2002}
}

@article{lande1998extinction,
  title={Extinction times in finite metapopulation models with stochastic local dynamics},
  author={Lande, Russell and Engen, Steinar and S{\ae}ther, Bernt-Erik},
  journal={Oikos},
  pages={383--389},
  year={1998},
  publisher={JSTOR}
}

@article{burkitt2006review,
  title={A review of the integrate-and-fire neuron model: I. Homogeneous synaptic input},
  author={Burkitt, Anthony N},
  journal={Biol. Cybern.},
  volume={95},
  number={1},
  pages={1--19},
  year={2006},
  publisher={Springer}
}

\end{document}